\documentclass{article}
\begin{document}
\title{Quantum Clock Synchronization: a Multi-Party Protocol} \author{Marko Kr\v{c}o and Prabasaj Paul\footnote{To whom correspondence should be addressed}\\Department of Physics and Astronomy\\Colgate University\\13 Oak Drive\\Hamilton NY 13346}
\maketitle
\begin{abstract}
We present a multi-party quantum clock synchronization protocol that utilizes shared prior entanglement and broadcast of classical information to synchronize spatially separated clocks. Notably, it is necessary only for any one party to publish classical information. Consequently, the efficacy of the method is independent of the relative location of the parties. The suggested protocol is robust and does not require precise sequencing of procedural steps.
\end{abstract}
\section{Introduction}
Clock synchronization has been the focus of intense study with the potential for both commercial and scientific rewards. Accurate timekeeping is a necessity for GPS satellites, long-baseline interferometry and the such. Currently used methods are based on two relativistic protocols by Einstein \cite{Ein}  and Eddington \cite{Edd}, but the physical limitations of these protocols have already been reached. These limitations are mostly due to the unstable dispersive properties of the intervening media \cite{Gio}. It is believed that new synchronization protocols using the peculiar properties of prior entanglement may provide increased robustness and accuracy. Several such protocols have been recently proposed. One two-party protocol proposed by Jozsa {\em et al.} \cite{Joz} (with related discussions in \cite{Joz2, Bur}) has the remarkable feature that the actual synchronization requires only classical communication (which does not carry any timing information) between the two parties, with no restriction on the mode of communication or the properties of the intervening medium. The protocol is implemented by supplying the parties with shared pairs of qubits in known maximally entangled energy eigenstates.

In this report, we extend and generalize the protocol proposed by Jozsa {\em et al.} to a multi-party version. The $n$ parties are initially supplied with shared $n$-qubit systems in known entangled energy eigenstates. (In general, no two of these qubits will be maximally entangled.) Each party measures the qubits in their possession in a certain predetermined basis. Next, the party in possession of the standard clock broadcasts their results (on the Internet, for instance). This public, classical information allows all parties to synchronize their clocks to the standard. Two features of the protocol are particularly noteworthy. First, maximal entanglement between pairs of qubits is not necessary for clock synchronization. Second, the protocol is symmetric and allows the parties to work independently; designation of a standard clock may be deferred until after all measurements have been made, and the public results may be accessed at the convenience of each party. 

In the following section, we present details of the protocol. We discuss some possible problems with the protocol and suggest solutions to them in the next section.

\section{The protocol}
We will now describe the $n$-party clock synchronization protocol in detail. There are $n$ spatially separated unsynchronized clocks, one of which is the standard clock. We designate the party in possession of the standard clock Alice, the publisher. The other clocks are in possession of receivers, one of whom is Bob. 

The qubits involved in the protocol have energy eigenstates $|0\rangle\equiv\left(\begin{array}{c}1\\0\\\end{array}\right)$ and $|1\rangle\equiv\left(\begin{array}{c}0\\1\\\end{array}\right)$ with energies $0$ and $\hbar\omega$, respectively. This defines the usual computational basis. We define the measurement basis as the set $|\pm\rangle\equiv\frac{1}{\sqrt{2}}\left(|0\rangle\pm|1\rangle\right)$.

We assume that the parties have access to an unlimited supply of identical distinguishable sets of $n$ non-interacting qubits, where each party is in possession of one qubit from each set. To ensure that the density matrix of the initial state is constant and known until measurements are made, the initial state of each set must be an energy eigenstate. As will be seen shortly, the initial state must also have non-zero (but not necessarily maximal) entanglement between each pair of qubits. Therefore, a suitable initial state for each $n$-qubit set is
\begin{equation}
\label{psi}
\vert\Psi\rangle=\frac{1}{\sqrt{n}} 
	\underbrace{	
		\left(
			\vert10\ldots00\rangle + 
			\vert01\ldots00\rangle +
			\cdots + 
			\vert00\ldots01\rangle
		\right)
	}_{n \:terms}
\end{equation}
where each term contains only a single qubit in the state $\vert1\rangle$. The symmetry explicitly built into $|\Psi\rangle$ is not necessary for the protocol to work; it has been done for ease of exposition. Note that the entanglement between pairs of qubits decreases with $n$ for this state.

Without loss of generality, we now focus our attention on Alice and Bob. The density matrix of the qubits in their possession is
\begin{equation}
\rho_{AB}=\frac{1}{n}
\left(\begin{array}{cccc}
n - 2 & 0 & 0 & 0 \\
0 & 1 & 1 & 0 \\
0 & 1 & 1 & 0 \\
0 & 0 & 0 & 0 \\
\end{array}
\right)
\end{equation}
in the computational basis and 
\begin{equation}
\label{rhoAB}
\rho_{AB}=\frac{1}{4n}
\left(\begin{array}{cccc}
n+2 & n-2 & n-2 & n-6 \\
n-2 & n-2 & n-2 & n-2 \\
n-2 & n-2 & n-2 & n-2 \\
n-6 & n-2 & n-2 & n+2 \\
\end{array}
\right)
\end{equation}
in the measurement basis $\{|++\rangle,|+-\rangle,|-+\rangle,|--\rangle\}$.

At standard time $t = 0$, Alice measures each qubit in her possession in the $\vert\pm\rangle$ basis. She then publishes the results of her measurement, labeling each qubit by its set. 
For each qubit measured as $|+\rangle$ by Alice, the corresponding qubit with Bob at standard time $t=0$ has the density matrix
\begin{equation}
\rho_{B}\left(t=0\right)=\frac{1}{2n}
\left(\begin{array}{cc}
n + 2 & n - 2 \\
n - 2 & n - 2\\ 
\end{array}
\right)
\end{equation}
in the measurement basis. This is just the (normalized) two-by-two upper left submatrix of $\rho_{AB}$ in (\ref{rhoAB}). To obtain $\rho_B(t)$, one may apply the time evolution operator to the density matrix above. (It is possible to do so unambiguously since the qubits are non-interacting.) This gives
\begin{equation}
\rho_{B}\left(t\right)=\frac{1}{2n}
\left(\begin{array}{cc}
n + 2\cos\omega t & n - 2 + 2i\sin\omega t\\
n - 2 - 2i\sin\omega t & n - 2\cos\omega t\\ 
\end{array}
\right).
\end{equation}

Bob, too, measures his qubits in the measurement basis at time $t_B=0$ by his clock (which is $t-t_B\equiv\Delta$ standard time). Focusing on the sets for which Alice measured $|+\rangle$, Bob will obtain the two possible outcomes with the following probabilities
\begin{equation}
\label{p}
P(\vert\pm\rangle)=\frac{1}{2} \pm \frac{\cos(w\Delta)}{n}.
\end{equation}
which are, of course, the diagonal elements of $\rho_B(t)$.
A parallel analysis of the sets for which Alice measured $|-\rangle$ yields similar results. Assuming that $|\omega\Delta|<2\pi$, the relative frequencies of the two results ($|+\rangle$ and $|-\rangle$) of his measurements will enable Bob to estimate $\Delta$ and adjust his clock accordingly.

Local measurements and a knowledge of Alice's published results has enabled Bob to synchronize his clock to the standard. Note that the measurements may be conducted independently by each party. Moreover, the designation of a standard clock may be undertaken after the measurements. Indeed, subgroups among the $n$ parties may choose different clocks as standards.
\section{Analysis}
An examination of our protocol reveals three possible problems. First, the qubits may acquire phases during transport to the spatially separated locations, which show up as relative phases between the terms in (\ref{psi}); this may, in principle, be eliminated by adiabatic transportation of the qubits. Second, the definition of the measurement basis may not be the same at each location; this is discussed in detail in \cite{Joz} and the two-frequency strategy suggested there to remedy this can be easily adapted to our protocol. Third, the accuracy to which $\Delta$ can be determined from a given number of sets decreases with $n$, since the amplitude of the variation of the probabilities in (\ref{p}) with time decreases with $n$. This can be attributed to the decrease in entanglement between pairs of qubits as $n$ increases. It is tempting to look for a different initial state than (\ref{psi}) that does not suffer from this problem. While bounds on entanglement in multi-qubit systems are currently under investigation, recent work \cite{Woo} suggests that the limits encountered here are universal.
\bibliography{qc_clock}  

\providecommand{\bysame}{\leavevmode\hbox to3em{\hrulefill}\thinspace}
\providecommand{\MR}{\relax\ifhmode\unskip\space\fi MR }
\providecommand{\MRhref}[2]{%
  \href{http://www.ams.org/mathscinet-getitem?mr=#1}{#2}
}
\providecommand{\href}[2]{#2}
\begin{thebibliography}{1}

\bibitem{Bur}
E.~A. Burt, C.~R. Ekstrom, and T.B. Swanson, Phys.\ Rev.\ Lett. \textbf{87}
  (2001), 129801.

\bibitem{Woo}
V.~Coffman, J.~Kundu, and W.K. Wootters.

\bibitem{Edd}
A.S. Eddington, \emph{The mathematical theory of relativity}, Cambridge
  University Press, 1924.

\bibitem{Ein}
A.~Einstein, Ann.\ D.\ Physik \textbf{17} (1905), 891.

\bibitem{Gio}
V.~Giovannetti, S.~Lloyd, L.~Maccone, and S.~M. Shahriar.

\bibitem{Joz}
R.~Jozsa, D.S. Abrams, J.P. Dowling, and C.P. Williams, Phys.\ Rev.\ Lett.
  \textbf{85} (2000), 2010.

\bibitem{Joz2}
\bysame, Phys.\ Rev.\ Lett. \textbf{87} (2001), 129802.

\end{thebibliography}
\end{document}